\documentclass[fleqn, 10pt]{wlscirep}
\usepackage[utf8]{inputenc}
\usepackage[normalem]{ulem}

\newcommand{\beginsupplement}{%
        \setcounter{table}{0}
        \renewcommand{\thetable}{S\arabic{table}}%
        \setcounter{figure}{0}
        \renewcommand{\thefigure}{S\arabic{figure}}%
     }

\title{Role of inter-hemispheric connections in functional brain networks}

\author[1,2*]{J. H. Mart\'inez}
\author[3,4]{J. M. Buld\'u}
\author[5]{D. Papo}
\author[1,6,7]{F. De Vico Fallani}
\author[7]{M. Chavez}

\affil[1]{INSERM-Institute du Cerveau et de la Moelle Épini\`ere. Paris, France}
\affil[2]{Grupo Interdisciplinar de Sistemas Complejos (GISC), Spain}
\affil[3]{Complex System Group \& GISC, Universidad Rey Juan Carlos, Madrid, Spain}
\affil[4]{Laboratory of Biological Networks, Center for Biomedical Technology, Universidad Polit\'ecnica de Madrid, Madrid, Spain}
\affil[5]{SCALab UMR CNRS 9193. Universit\'e de Lille 3, France}
\affil[6]{INRIA Paris, UPMC Univ. Paris 06, Paris, France}
\affil[7]{CNRS-UMR 7225, H\^opital Piti\'e-Salpetri\`ere, Paris, France}

\affil[*]{Corresponding author: johemart@gmail.com}

\keywords{Brain networks, hemispherical balance, complex networks, competition, robustness}

\begin{abstract}
Today the human brain can be modeled as a graph where nodes represent different regions and links stand for statistical interactions between their activities as recorded by different neuroimaging techniques. Empirical studies have lead to the hypothesis that brain functions rely on the coordination of a scattered mosaic of functionally specialized brain regions (modules or sub-networks), forming a web-like structure of coordinated assemblies (a network of networks). The study of brain dynamics would therefore benefit from an inspection of how functional sub-networks interact between them. In this paper, we model the brain as an interconnected system composed of two specific sub-networks, the left ($L$) and right ($R$) hemispheres, which compete with each other for centrality, a topological measure of importance in a networked system. Specifically, we considered functional brain networks derived from high-density electroencephalographic (EEG) recordings and investigated how node centrality is shaped by interhemispheric connections. Our results show that the distribution of centrality strongly depends on the number of functional connections between hemispheres and the way these connections are distributed. Additionally, we investigated the consequences of node failure on hemispherical centrality, and showed how the abundance of inter-hemispheric links favors the functional balance of centrality distribution between the hemispheres.
\end{abstract}

\begin{document}
\flushbottom
\maketitle
\section*{Introduction}
Like almost all real networked systems, both structural and functional brain networks have been found to display an heterogeneous structure\cite{Eguiluz2005,Bullmore2009,Hayasaka2010,VandenHeuvel2011}. Whereas in random networks nodes have approximately the same number of links, in heterogeneous ones nodes are unevenly connected to each other, and some of them, called hubs, may have a very large number of links. Far from being a mere statistical fact, heterogeneity plays an important role in a variety of dynamical processes such as diffusion, information transmission, network vulnerability, control and synchronization \cite{newman2010}. Several studies have focused on the role of hubs within and between brain networks \cite{Hagmann2008b}. The presence of hubs in brain networks has been related to a reduction of wiring costs, as hubs behave as integrator and distributor of information through the network in combination with few long-range connections to other brain modules \cite{Achard2006,VandenHeuvel2011}. Hub failure, quantified in terms of loss of connectivity, has been associated with the emergence of various brain pathologies \cite{Bassett2008,Navas2015}, while hub's importance has been shown to be altered with aging \cite{Meunier2009a}. Although hubs seem to play an important functional role in networks, characterization and quantification of their functional role from experimental data is not trivial \cite{Clauset2009}. The simplest measure of a node's importance is given by the total number of its connections (node's degree)  or weights (node's strength). However, hubs cannot be ranked using local metrics (degree or strength) as they fail to account for the role of the network into which the node is embedded. To address this issue, various global metrics have been proposed, including \textit{closeness} \cite{newman2010}, \textit{global efficiency} \cite{Latora2001}, \textit{node betweenness} \cite{Freeman1977} or \textit{eigenvector centrality}\cite{newman2010}. At a global level, brain regions tend to connect to regions with a similar number of connections \cite{Eguiluz2005}, which naturally leads to the formation of a rich club \cite{Colizza2006} of tightly connected hubs \cite{VandenHeuvel2011}. A further difficulty in node ranking arises from the existence of a mesoscopic topological structure intermediate between the macroscale of the whole network and the microscale of individual nodes. At mesoscopic scales, both anatomical and functional networks are characterized by a modular structure \cite{Meunier2009a,Fortunato2010}. These mesoscale structures or communities can be used to better characterize the role of important nodes, allowing to classify them as local and global hubs based on connectivity within and without the community a given node belongs to \cite{guimera2005}.

Hub ranking metrics are affected not just by the way that these nodes interact with each other but also by the mere existence of community structure. For example, in a modular or a network-of-networks structure the interplay between structural and dynamical properties such as vulnerability \cite{Gao2011} or synchronization \cite{Aguirre2014} behave in a counter-intuitive way due to the existence of mesoscale structure. More recently, it was shown that centrality distribution of network modules turns out to strongly depends on the connections between clusters\cite{Aguirre2013}. On the one hand, the specific nodes connecting network modules influence the centrality of the modules themselves; on the other hand, node centrality strongly depends on the community the node belongs to. As a consequence, the connectivity between modules may range between two extreme strategies, one in which the modules' central nodes connect to each other (central-central connection or CC), and another in which the lowest-centrality nodes are connected (peripheral-peripheral connection or PP).  As shown in \cite{Aguirre2013},  PP connections between network modules favour the centrality retained by strongly connected modules (i.e., those with more within-module connections), while a CC connection strategy benefits weakly connected nodes.

Within this framework, functional brain units at various scales can be thought of as modular networks or networks-of-networks, as they continuously interact in various ways, e.g. in a synergistic or antagonistic or otherwise modulatory manner. By far the most studied form of modularity in the brain is that represented by the structural and functional hemispheric subdivision \cite{Sperry1969}. An early but still cited model of interhemispheric cross-talk \cite{Kinsbourne1974} proposed that the brain is a highly reciprocally interconnected neural network, characterized by a constantly shifting pattern of locally higher and lower activation levels. Studies using fMRI suggest that hemispherical interactions are characterized by both functional cooperation and competition \cite{Fornito2012,Cocchi2013,Doron2012}, and patterns of correlated brain activity at rest can be used to evaluate the extent of hemispheric asymmetry \cite{Agcaoglu2015}. However, the exact functional mechanisms through which the hemispheres interact are still not well understood\cite{Agcaoglu2015,Nielsen2013}. Here, we address two outstanding questions: to what extent are brain hemispheres engaged in a competitive behaviour in terms of nodal centrality? and, how does the importance of each hemisphere rely on interhemispheric connections?

To do so, we evaluate how the centrality of resting-state functional brain networks of healthy individuals is distributed between brain hemispheres and how functional inter-hemispherical links influence this distribution. We use eigenvector centrality as an indicator of node importance \cite{Navas2015} due to the fact that it takes into account the whole connectivity of the network and not only the connectivity of each node. Eigenvector centrality can be derived in a rather straightforward way from the eigenvector associated with the largest eigenvalue of the (weighted) network adjacency matrix \cite{newman2010}. The eigenvector centrality of a given network is then simply the sum of the centrality of all its nodes \cite{Aguirre2013}. The allocation of centrality between hemispheres was quantified at different frequency bands and for two different resting conditions (closed and open eyes). We also studied the robustness of hemispheric centrality against failure of individual nodes, as well as a function of the number of inter-hemispheric links.

\section*{Results}
\subsection*{Node centrality}
We analyzed the distribution of centrality of the functional brain networks of a group of $54$ healthy individuals at rest in open and closed eyes conditions. Brain networks were obtained, at the sensor level, from EEG recordings containing $N=48$ electrodes (24 per hemisphere). Link weights were quantified using the imaginary part of the coherence between time series of every pair of sensors (See Section Methods for a detailed description of the full process for obtaining the functional networks).
In the following, we considered each brain hemisphere as a sub-network. 

\begin{figure}[ht]
 \centering
 \includegraphics[width=0.9\textwidth]{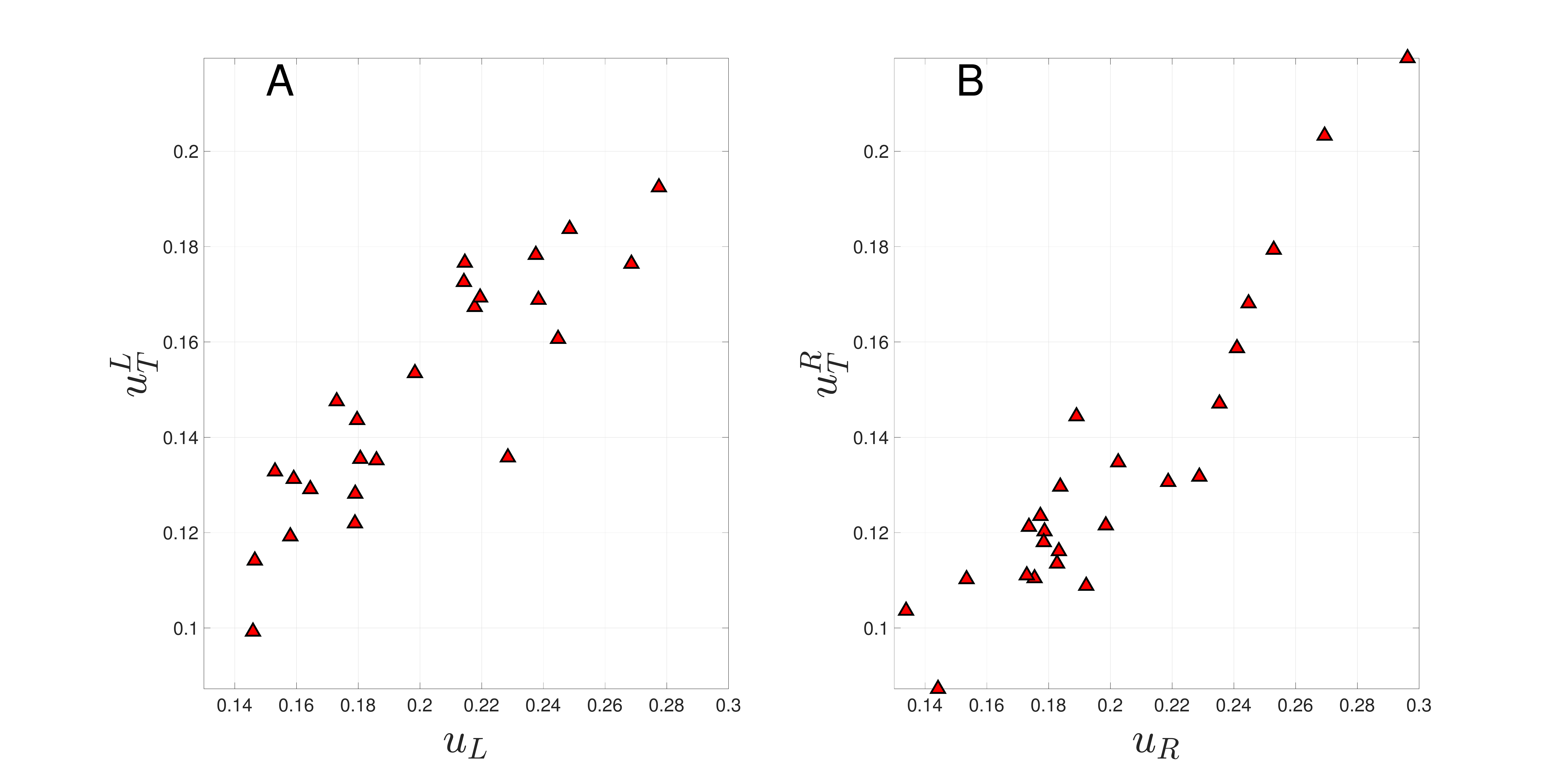}
 \caption{
 Left (\textbf{A}), and Right (\textbf{B}) hemispheres. Global centrality (vertical axes) is obtained from the complete matrix $T$, when all functional connections between hemispheres are maintained. Local centrality is extracted from the fully connected sub-networks $L$ and $R$ when hemispheres are ``functionally" disconnected. Results for the rest of the frequency bands and for the EO condition indicate a similar behaviour (see Fig. S1,
 Supplementary information). 
}\label{fig:01}
\end{figure}

We computed node centrality for subjects under two experimental conditions: resting-state with eyes open (EO) and eyes closed (EC), differentiating the hemisphere the nodes belong to. We distinguished between the centrality over the whole brain (i.e. within the NoN) and the centrality within one hemisphere (i.e., within each sub-network). The global centrality $u_T$ is obtained as the eigenvector of the adjacency matrix $T$, a fully connected graph of $N=N_L+N_R=48$ nodes associated with the ensemble of two connected hemispheres, where $L$ and $R$ stand for left and right hemispheres, respectively. Local centralities $u_{L,R}$ are the eigenvectors associated to the largest eigenvalues of adjacency matrices $L$ and $R$ when both hemispheres are disconnected. When comparing local and global centrality of the nodes one may expect that hubs of each hemisphere to also be the hubs when both hemispheres are connected.
This is what we observe in Fig. \ref{fig:01}, where we show the interplay between global and local centrality measures $u_{L,R}$, vs $u_T^{L,R}$ for both hemispheres in the EC condition, in the $\alpha$ band. In Fig. \ref{fig:01}\textbf{A}(\textbf{B}), global centrality for left (right) hemisphere is defined as $u_{T}^L = u_T\{1, ..., N_L\}$ ($u_{T}^R = u_T\{N_L+1, ..., N\}$) and local centrality for $L$ ($R$) is defined as $u_L$ ($u_R$). This figure reveals a positive trend between local centrality $u_{L, R}$ and global centrality $u_{T}$ for both hemispheres. We specifically focused on $\alpha$ band activity in the EC condition, where resting activity was associated with higher activation. However, a similar behaviour was found for the rest of frequency bands, as well as for the EO condition (see Fig. S1 
of Supplementary Information for details).

These results suggest that the existence of functional inter-hemispherical connections does not introduce significant changes in node centrality when compared with the centrality obtained for each separate hemisphere.

\subsection*{Evaluating the centrality distribution between hemispheres}
We further investigated the centrality distribution at a macroscale level. Since the total centrality of the NoN can be normalized to one, the centrality distribution can be interpreted as a competition process, e.g., if the left hemisphere retains a global centrality $C_L$, the right one has $1-C_L$ (see Methods). 

\begin{figure}[ht]
 \centering
 \includegraphics[width=0.8\textwidth]{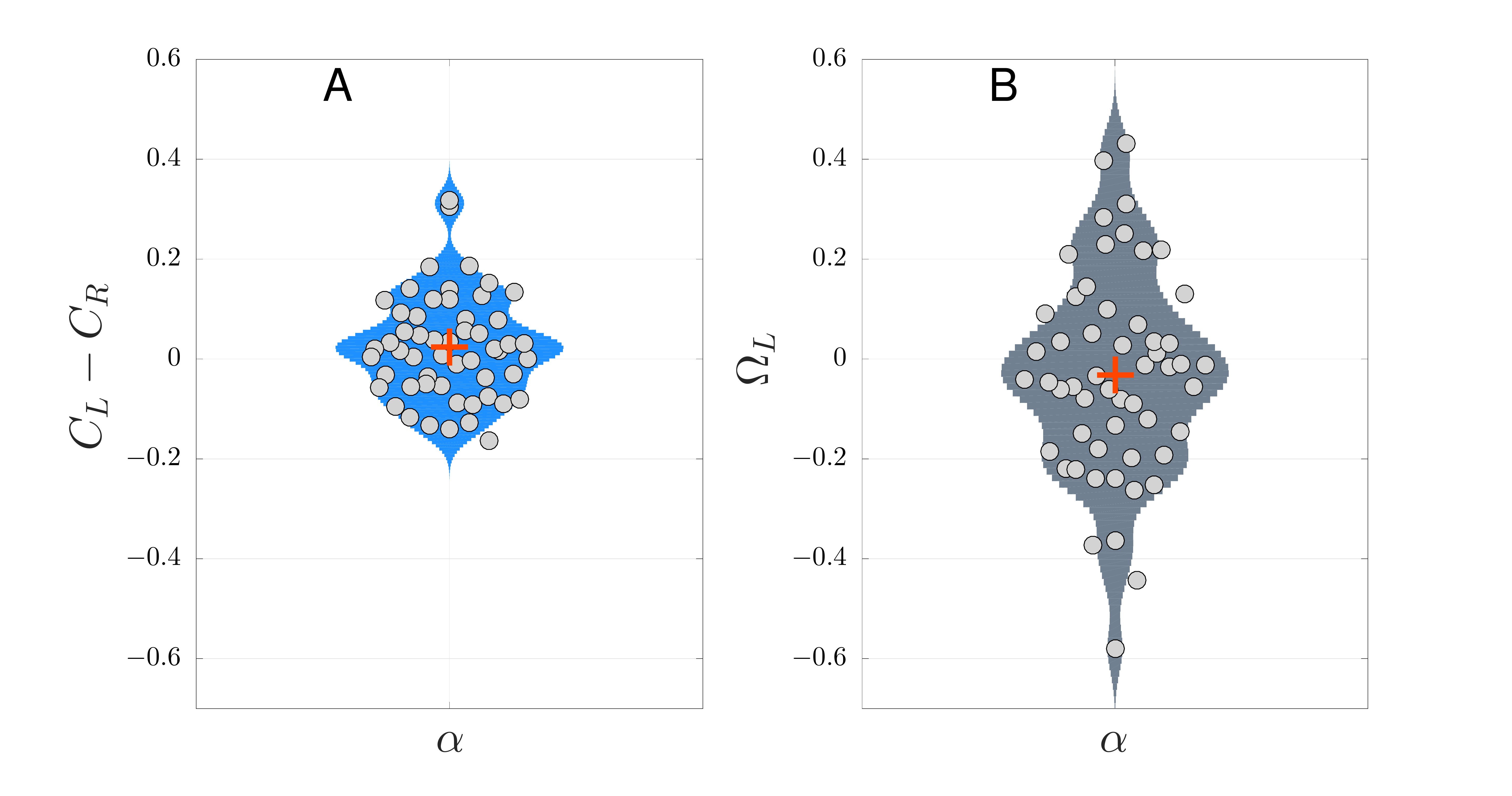}
 \caption{
\textbf{A}. Differential centralities ($C_L - C_R$) along 54 subjects, for the $\alpha$ band. \textbf{B}. Competition Parameter respect left hemisphere $\Omega_L$ for all subjects. In both subplots, red cross indicate the mean values.}\label{fig:02}
\end{figure}

First, we calculated how the centrality of the whole NoN was distributed along the two hemispheres by computing the difference of hemispherical centrality ($C_L -C_R$), for each subject. For simplicity, we chose the left hemisphere as the reference network. In this way, we obtained the centrality difference, henceforth termed {\it centrality contrast} ($C_L - C_R$) for the 54 subjects. Interestingly, results show that the centrality accumulated by $L$ or $R$ is quite similar to each other. Fig. \ref{fig:02}\textbf{A} shows how the difference of centrality in EC was close to zero (Mann-Whitney U test, $p>0.7$), indicating a situation of approximate balance in the centrality distribution between hemispheres. This trade-off pattern appears in all bands even in EO condition (see Fig. S2\textbf{A} and table S1
in Supplementary information). 

While no significant differences were found when comparing fully connected hemispherical networks, a question arises as to the role of the inter-hemispherical links: do they promote the centrality of one hemisphere or, on the contrary, do they promote a balance? To clarify this issue, we computed the competition parameter with respect to the left hemisphere $\Omega_L$,  an indicator of how the inter-hemispherical links may favor one or rather the other hemisphere. To compute $\Omega_L$, we first estimate the highest/lowest possible centrality of each hemisphere may reach by rewiring the inter-hemispherical connections. In this context, the rewiring implies reshuffling all elements within the inter-hemispherical adjacency matrix $P$ that connects both $L$ ad $R$. In the absence of theoretical bounds, these rewirings are necessary to assess extreme values of $\Omega_{L,R}$ as a function of inter-hemispherical links \cite{Aguirre2013}. An illustrative example of how this rewiring algorithm works is shown in the Supp. Info. (Fig. S3).

Figure \ref{fig:02}\textbf{B} shows the $\Omega_L$ distribution of the $\alpha$-EC condition for the $54$ individuals. Similar with the centrality differences, we performed the previous statistical test between both condition in all bands and the null hypothesis with sensitivity of $0.05$ was not rejected. In other words, in fully connected networks, the competition parameter tends to zero, hemispheres do not compete for centrality and remain in functional balance. Results obtained in other bands for EC and EO (see Fig.S2\textbf{B} and table S1 of Supp. Info.

Taken together, these results suggest that when working with fully connected networks there is no major distinctions between the centrality of brain hemispheres either at micro (node) or macro (network) scales. This would indicate that hemispheres are close to a functional balance with regard to centrality distribution. Nevertheless, it is crucial to investigate to what extent this balance is a consequence of the particular network topology: Is the type of network-of-networks topology responsible of the balanced division of the network centrality?

\subsection*{Inter-hemispherical links and the centrality competition}

All previous results considered fully connected networks, i.e. there existed a connection between all pair of nodes with a value greater than zero. However, functional networks are usually thresholded, maintaining only those relevant links of the network \cite{Papo2014,Papo2015}. Therefore, we are going to evaluate the effect of thresholding the functional network and re-evaluate the effect of the inter-hemispheric links on the  centrality distribution between hemispheres. 
The reason behind is that a recent theoretical study has demonstrated the existence of different regimes in the centrality distribution according to the number of inter-links between networks \cite{Aguirre2013}.

In this way, we keep the strongest links of $L$ ($R$) and disregard all weaker links such that each hemisphere remains as a single connected component. Note that, following this procedure may lead to both hemispheres to have different thresholds. Nevertheless, we select the lowest threshold of the two hemispheres in order to guarantee that all nodes in each sub-network are connected. Then, we repeat this procedure for the functional network of each subject.

\begin{figure}[ht]
 \centering
 \includegraphics[width=1.0\textwidth]{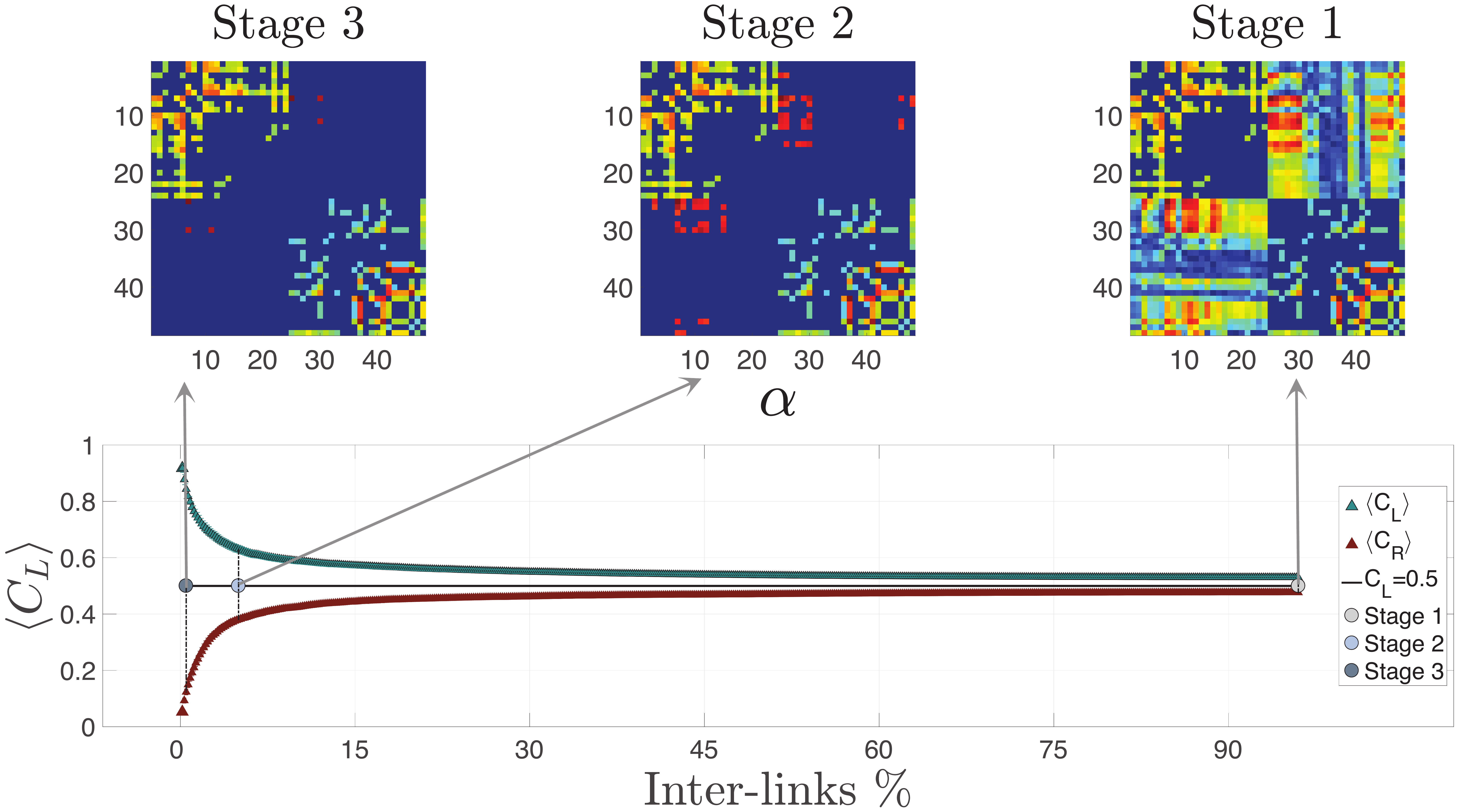}
 \caption{
Connectivity matrices from a given subject in two regions according to the number of  inter-hemispherical links in the matrix $P$. $L$ and $R$ hemispheres are thresholded such that the corresponding sub-networks have a unique connected component. \textbf{Stage 3} contains only three weighted inter-links (i.e., $\sim0.5\%$ of all possible inter-links); \textbf{Stage 2} is at the border between the two regions and it contains thirty inter-links ($\sim5.0\%$); \textbf{Stage 1} contains all $P$ elements, i.e., all inter-hemispheric links (100\%). \textbf{Bottom panel}. Evolution of $\langle C_L \rangle$ and $\langle C_R \rangle$ as a function of the number of inter-hemispherical links (note that $\langle C_R \rangle = 1- \langle C_L \rangle$ ). Red triangles correspond to right-dominant individuals and green ones to left-dominant ones. The three dots in the horizontal axis simply indiacte the three stages of connectivity matrices in the upper panel.}\label{fig:03}
\end{figure}

The sparse sub-networks corresponding to the hemispheres were connected by adding their inter-hemispherical links. In order to investigate the effect of the inter-hemispherical links in detail, we included links one-by-one in the inter-hemispherical block matrix $P$, beginning from the strongest link. We then computed the associated centrality distribution (i.e., $C_L$ and $C_R$) and included a new link (the following in strength) until all inter-hemispherical links were re-introduced. We could thus quantify  the values of $C_L$ ($C_R$) as a function of percentage of inter-hemispherical links. 
Figure \ref{fig:03} \textbf{Bottom panel} shows the evolution of the $\langle C_L \rangle$ and $\langle C_L \rangle$ for all subjects in $\alpha $-EC when inter-hemispheric links are added once at time (see Fig. S4
of Supp. Info. for all bands in both conditions).
Results reveal a clear monotonic behavior of the hemispheric centrality, i.e., for the hemisphere with higher centrality, the hemispheric centrality decreases as the number of inter-networks links is increased, while the behaviour is the opposite for the hemisphere with lower centrality.

The majority of studies investigating the role of nodes in functional networks, specifically in the characterization of functional hubs, typically tend to report significant results for sparse networks rather than strongly connected ones? Our results show a crucial dependency of hemispherical centrality on the amount of inter-hemispheric links that coordinate the activity between them, leading to two different regimes: (1)  A regime where the number of inter-hemispherical connections is low (from $\sim0.5\%$ to $\sim5.0\%$ of the inter-hemispheric links) and the centrality is hoarded by one hemisphere and, (2) a region where the number of inter-links is high and centrality is evenly distributed between the two hemispheres. 

We can observe how the number of inter-links determines the balance of centrality between hemispheres, a fact that must be taken into account when computing not only hemisphere centrality but also the centrality of each node of the functional network and of the hubs in particular. If we take into account that studies investigating the role of the nodes in functional networks, specifically in the characterization of functional hubs, are based on threshold values that typically ranks bellow 10\% of strongest links \cite{Papo2014b,Bullmore2009}, it is clear that this effect must be taken into account when interpreting the results.

\subsection*{Hemispherical Robustness against node failure}
So far we illustrated our results on centrality distribution under normal conditions (i.e., EC and EO, resting state) but no information was provided on the effect of network dysfunction. To further investigate the role of each node in the centrality distribution of the whole network, we studied the effects of node removal, a common way of modeling the effect of brain lesions which can be used to assess the robustness of brain networks \cite{stam2009a,Stam2007,Kaiser2007,Alexander-Bloch2010}.

Insofar as centrality distribution depends on interhemispheric links, robustness evaluation should also be inspected by considering the impact of sele removal (see caption of Fig. \ref{fig:03} for details) on hemispherical centrality.  Specifically, we removed a node $i$ from the left (right) hemisphere and computed the new hemispherical centrality $C^L_*$ ($C^R_*$) without that node. Next, we estimated the percentage of network damage as the difference with the actual centrality of that hemisphere $C^L$ ($C^R$). The local impact, here defined for left as reference, $l_{imp}^L(i)$ is defined as the percentage $l_{imp}^L(i)=\frac{(C^L_*-C^L)}{C^L}\times 100$. At the same time, we measured the local contribution $lc^L(i)$ of the node $i$ being removed as the percentage of importance inside its hemisphere, $lc^L(i)=\frac{u^L_T(i)}{\sum_i^{N_L} u^L_T(i)}\times 100$. This allowed quantifying how the relevance or the importance of a node inside each hemisphere is related to the modifications of the centrality acquired by its own hemisphere.

\begin{figure}[hbtp]
 \centering
 \includegraphics[width=0.8\textwidth]{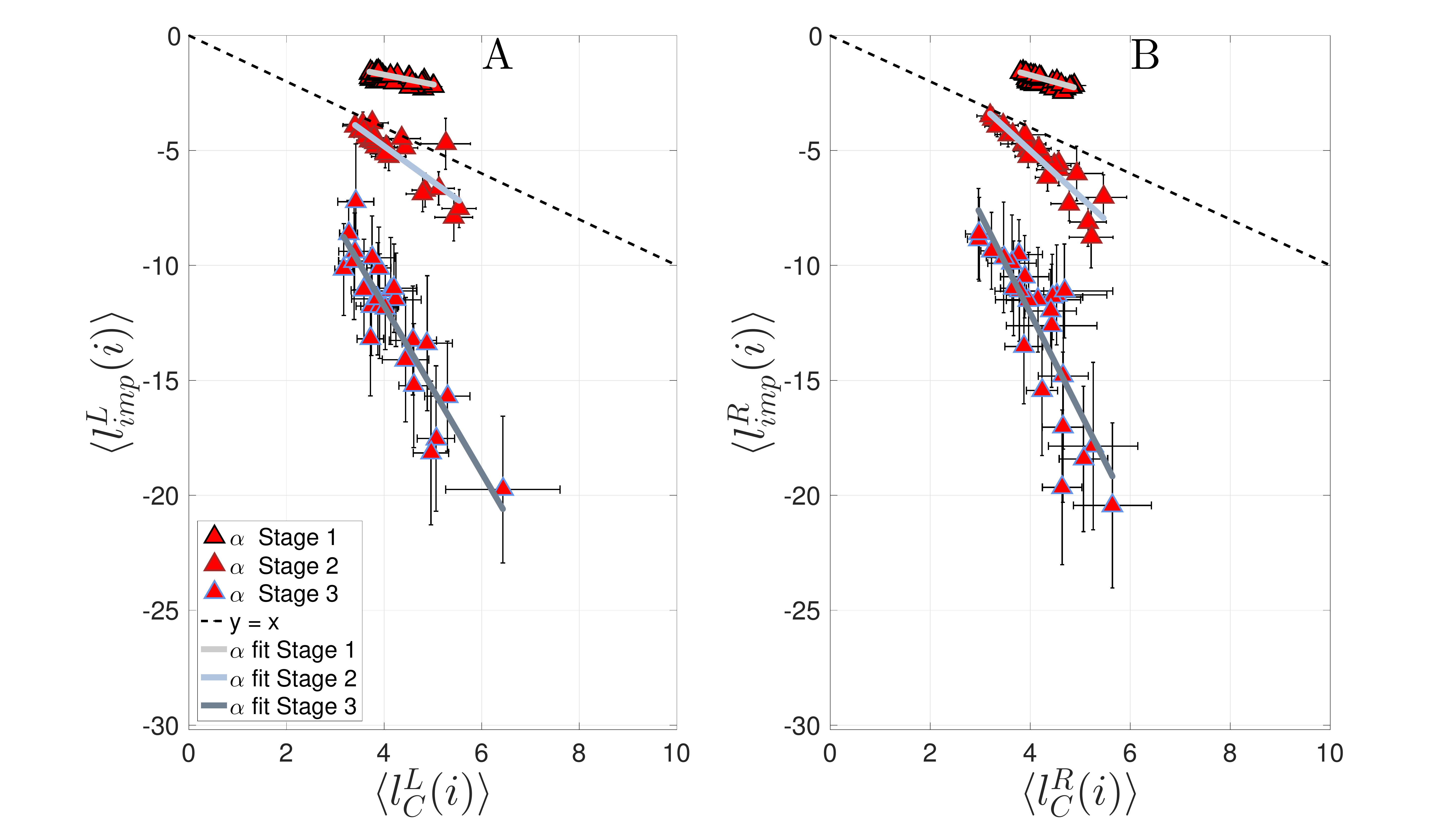}
 \caption{
 In all plots, triangles correspond to the average values from all subjects. Bars represent the standard error for impacts (horizontal) and centrality contributions (vertical).
 Subplot \textbf{A} (\textbf{B}) refers to the centrality impact in the left (right) hemisphere.
In both panels, groups of nodes from Stage 1 correspond to the upper cloud of triangles, Stage 2 to the middlle cloud and Stage 3 to the bottom group. The black dashed line is just a reference and corresponds to $\langle l_{imp}^L(i)\rangle=\langle lc^L(i)\rangle$, where the percentage of impact in the hemisphere centrality is the same as the percentage of importance of the removed node. For each stage, a solid line allows to better follow the linear trend between local centrality contribution and local impact.}\label{fig:04}
\end{figure}

Plots in Fig. \ref{fig:04} show the behavior of $\langle l_{imp}^L(i)\rangle$ vs. $\langle lc^L(i)\rangle$ in the stages defined by the number of inter-hemispherical links for both left and right hemispheres. Figure \ref{fig:04} allows to evaluate whether the impact of a node is higher/lower than its importance or, in other words, to see if a node with importance $x$ has an impact higher or lower than $x$. The slope of the linear fitting in each stage indicates a damage ratio respect to $\langle l_{imp}^L(i)\rangle = \langle lc^L(i)\rangle$ , and also reflects how important a failure is.

In all plots of Fig. \ref{fig:04}, the Stage 1 (i.e., when all inter-hemispherical links are maintained) corresponds to the group of nodes lying above the dashed line given by ($\langle l_{imp}^L(i)\rangle=\langle lc^L(i)\rangle$). This indicates that the impact on the local centrality is always lower than the importance of the node itself. Regardless the removed node in a network, a node never entails an important impact in the centrality distribution. Note that the damage, referred in the vertical axes, would be always lower than 5\% no matter what node fails at Stage 1. In both conditions (EC and OE), the slopes for left and right hemispheres are lower than one (see Tab \ref{tab:tab03}). The explanation of this behavior is that once a node is removed, the centrality distribution is mainly maintained by the remaining nodes of the same hemisphere, despite part of the centrality is captured by the opposite hemisphere. This phenomenon promotes the resilience of the hemisphere against random failures in Stage 1 and also maintains the functional network-of-networks close to a centrality balance.

For the connectivity obtained in the Stage 2 of Fig. \ref{fig:03} (the one maintaining the 5.0\% of the links) results are localized just below the dashed line in Fig. \ref{fig:04}. Although the local importance of nodes is maintained, we can observe how the damage in the centrality accumulated by hemispheres is increased when compared to the previous stage, and how the local impact of the nodes is close to its local contribution. In this way, the lower the number of inter-hemispherical links, the farther the system is from the balance in centrality distribution (and the higher the centrality captured by one of the hemispheres). The fact that centrality distribution of neworks-of-networks is very sensitive to the inter-links when their number is moderate to low was previously reported in \cite{Aguirre2013}, where it was shown that when the connectivity between two networks is sparse, the centrality captured by the dominant network increases significantly. In that sense, our results go one step further, as we observe that the leak of centrality towards other networks as a node fails is also promoted when links connecting networks are scarce.
 In addition, while the linear tendency between node contribution and impact is retained, we observe an increment of the slope, indicating that the role of the removed node is now taking more importance. In other words, removing central nodes now induces more damage in network centrality than removing peripheral ones.

\begin{table}[h]
\centering
\begin{tabular}{llcrlcrlcrlcrl}
\hline
\noalign{\smallskip}
& \multicolumn{3}{c}{Stage 3} & \multicolumn{3}{c}{Stage 2} & \multicolumn{3}{c}{Stage 1} \\

\noalign{\smallskip}
\cline{3-4}
\cline{6-7}
\cline{9-10}
\noalign{\smallskip}
Band & & $m_{EC_L}$ & $m_{EC_R}$ & & $m_{EC_L}$ & $m_{EC_R}$ & & $m_{EC_L}$ & $m_{EC_R}$ & \\
\noalign{\smallskip}
\bottomrule[1.2pt]

\noalign{\smallskip}
$\mathbf{\alpha}$ && 
$-3.629$ &  $-4.322$ && $-1.541$ &  $-2.00$ && $-0.421$ &  $-0.608$ & \\
\hline
\end{tabular}
\caption{
Slopes of linear fits of $\langle l_{imp}^L\rangle$ vs. $\langle l_c^L\rangle$ from Fig. \ref{fig:04}, for $\alpha$ band and EC condition in all stages. Note the increase of the (negative) slope as well as the system moves from stage 1 to stage 3.}\label{tab:tab03}
\end{table}

Finally, Stage 3 (0.5\% of the inter-hemispheric links are maintained) corresponds to the cloud of triangles observed at the lowest part of panels in Fig. \ref{fig:04}. Under this configuration, the impact of node failure is dramatically increased when compared with the previous stages: damage to the hemispherical centrality is more than three times the local contribution of a node. Thus, removing a node with a local centrality of 5\% would lead to a loss in centrality of more than 15\%. 
In addition, the slope of the fitting is the highest, indicating that differences between removing network hubs or peripheral nodes increase even further.
These results show that the lower the number of inter-connections between hemispheres, the highest the centrality vulnerability of the hemispheres. Importantly, these conclusions are independent of the considered hemisphere, the condition under study (CE or OE) or the frequency band being analyzed (See Fig. S5, Tabs. S2 and S3
of Supp. Info. for a detailed study).

\begin{figure}[ht]
 \centering
 \includegraphics[width=1.0\textwidth]{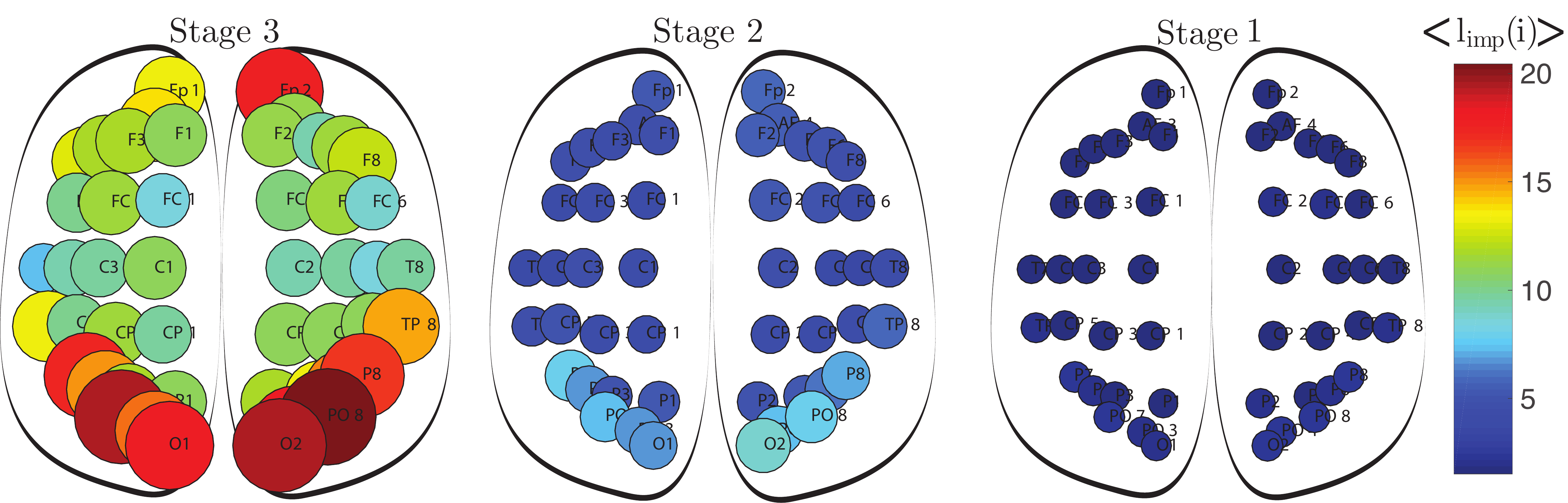}
 \caption{
Average local impact in $\alpha$-EC condition at the three stages: (i) Stage 1, all inter-hemispheric links, (ii) Stage 2, 5.0\% of the inter-hemispheric links and (iii) Stage 3, only 0.5\% of the inter-hemispheric links are considered. The radius and color of each node is proportional to the average local impact. Note how the local impact of the nodes increases as we move from Stage 1 to Stage 3 emerging the occipital lobe as the region with more vulnerable regions.}\label{fig:05}
\end{figure}

Finally, 
we plot in Fig. \ref{fig:05} the spatial distribution of the nodes, with size and color indicating the percentage of local impact in the $\alpha$-EC condition. We observe that when all inter-hemispheric links are considered, the local impact of the nodes is low and similar regardless their local centrality (Fig. \ref{fig:05}, \textbf{Stage 1}). On the contrary, when only few inter-hemispherical links are considered (Fig. \ref{fig:05}, \textbf{Stage 3}), the local impact of all nodes drastically increases, as well as certain heterogeneity emerges. Interestingly, vulnerable nodes are enclosed in a region located in the parieto-occipital lobe (Fig. S6
 of Supp. Info. shows the comparison respect EO condition).

Results strongly suggest that inter-hemispherical links are crucial to understanding local damages in a functional network-of-networks as a function of centrality distribution. Nevertheless, this point of view has been traditionally disregarded when analyzing functional brain networks. In fact large number of past studies have focused on other network parameters, such as the clustering coefficient and average shortest path. Here, we assess the dependence of these two network parameters 
on the inter-hemispherical links. To do this, we calculated the impact of node failure on the clustering coefficient $l_{imp_c}(i)$ and the shortest path $l_{imp_d}(i)$. As in the previous results, we compared these impacts with respect to the importance of node removal (i.e., the local contribution $l_{c}$). Contrary to centrality distribution, no significant changes could be observed for the clustering coefficient or the shortest path  (see Fig. S7 and Fig S8
of Supp. Info. for a detailed study). Both clustering and shortest path remained almost unaltered when the number of inter-hemispheric links was modified.

\section*{Discussion}
We addressed the issue of centrality distribution when a functional brain network is considered as a network-of-networks and examined how the interconnections between brain hemispheres determine a functional balance in terms of centrality distribution. We also evaluated the network robustness and identify the localization of regions that suffer the most due to node removal.

Our results show that the number of inter-hemispheric connections leads to two different scenarios for the distribution of centrality. When all functional connections between hemispheres are considered, centrality appears to be evenly distributed between the hemispheres.  On the other hand, for sparse networks, i.e., when only the strongest inter-hemispheric connections are considered, the centrality distribution becomes strongly unbalanced. In this scenario, the hemisphere with higher $\lambda_1$ (i.e., spectral radius), concentrates a high percentage of the overall centrality. As a consequence, the leading nodes of the weak hemisphere are strongly downgraded in the centrality ranking, hiding their importance even in their own community. Importantly, classical studies on the detection of network hubs, commonly consider a rather low number of connections and therefore run the risk of overlooking the existence of leading nodes in the weak hemisphere.

The three aforementioned scenarios are crucial to understand the re-distribution of centrality as nodes are removed from one hemisphere. When the number of interconnections is high, the damage of an hemisphere is always lower than the importance of its removed nodes. This result indicates that centrality is mainly redistributed among the nodes belonging to the community (hemisphere) of the removed node. On the contrary, when the number of inter-links is low, hemispheres are less robust to damage: the loss of centrality suffered by the hemisphere always exceeds the centrality of the removed node, as centrality redistributes towards the other hemisphere.

When all inter-hemispherical connections are considered, the results point to a state of functional balance that, in the context of centrality, is merely a consequence of the hyperconnectivity between hemispheres. This functional balance is also captured by the competition parameter $\Omega$, which is close to zero. On the other hand, the existence of both scenarios cannot be detected when only clustering or shortest path analysis are considered, which suggests that both measures are not able to capture the organization of functional networks as networks-of-networks.
 
Interestingly, the cortical regions that lead to the most severe deterioration of the hemispherical centrality are located in the parieto-occipital lobe for all conditions. This is in line with the role of parieto-occipital areas in the alpha band power modulation that has been associated with automatic gathering of non-specific information resulting from more interactions within the visual system \cite{Yan2009,Marx2004}

\section*{Methods}
\subsection*{Experimental Setup}
Database consists of 54 healthy subjects recorded in two different baseline conditions; 1-minute eyes closed (EC) and 1-minute eyes opened (EO), both at resting state. All subjects gave written informed consent for participation in the study, which was approved by the local ethics committee CPP-IDF-VI of Paris (no 2016-A00626-45). All experiments were performed in accordance with relevant guidelines and regulations. Subjects were comfortably seated on a reclining chair in a dimly lit room. During EO they were asked to avoid ocular blinks in order to reduce signal contamination.  The EEG data were recorded on a commercial system (Brainproduct GmbH, Munich, Germany) with a sampling rate of $200$~Hz. All the EEG signals were referenced to the mean signal recorded from the ear lobes. Data were subsequently down-sampled to $100$~Hz after applying a proper anti-aliasing low-pass filter. The electrode positions on the scalp followed the standard 10-10 montage \cite{Srinivasan2012}. 5 electrodes are excluded and only 56 are retained for the subsequent analysis. From the previous, 8 electrodes associated to the sagittal plane were removed so as to differentiate between two hemispheres. This way, the final amount of nodes used in this study is therefore 24 electrodes per hemisphere.

\subsection*{Imaginary Coherence} 
EEG signals arise from tangential and radial oriented cortical sources respect to the scalp surfaces, with many different electric conductivities (i.e., skull, hair, skin,...) which can influence the measurement of these sources, blurring the real EEG acquired data. This may lead to spurious correlation estimates due to the volume conduction effects. To overcome this issue, the imaginary coherence ($iCoh_{i,j}(f)$) has been proposed as a satisfying method to assess brain connectivity based on the frequencies of the brain signals \cite{Nolte2004a}. This measure of functional connectivity has been proved to be effective to avoid the field spread and cross-talk residual effects in EEG data \cite{GarciaDominguez2013a,DeVicoFallani2012}. Given two zero-mean time series $x(t)$ and $y(t)$ for channels X and Y respectively we can compute their complex Fourier transforms $S_x(t,f)$ and $S_y(t,f)$ \cite{Nolte2004a}. Then we obtain the cross spectrum as $S_{X,Y}(t,f)= \langle S_X(t,f) \cdot S^*_Y(t,f)\rangle$, where $\langle \cdot \rangle$ is the expectation operator. Hence the imaginary coherence $iCoh_{X,Y}(f)$ is defined as the imaginary part of the normalized cross spectrum:
\begin{equation}
iCoh_{X,Y}(f)=Im\{\frac{S_{X,Y}(f)}{\sqrt{S_{X,X}(f) \cdot S_{Y,Y}(f)}}\}
\end{equation}
\subsection*{Functional Network Construction: Hemispheres and Interlinks}
The imaginary coherence gives us a value that can be used as a {\it weight of communication} between two brain sites. In this way, we use the imaginary coherence to obtain a $N \times N$ symmetric matrix (with $N=48$), which can be interpreted as a weighted adjacency representation of a graph. For each subject, we obtained four weighted averaged networks for  the following frequency bands: $[4\leq \theta\leq 7]$,  $[8\leq\alpha\leq 13]$, $[14\leq \beta\leq 29]$ and $[30\leq \gamma\leq 40]$~Hz, in both conditions EC and EO.

 Initially, we use the \textit{fully connected networks} to maintain as many information as possible contained in the links' weights. Networks associated to each individual contain at the same time the information about two sub-networks $L$ and $R$, corresponding to the left and right hemispheres, respectively. Each of these two modules has $N_L=N_R=24$ nodes. Finally, we define the matrix $P$ as the inter-hemispherical block matrix containing only the interlinks that connect both hemispheres. In this way, the structure of the whole brain network-of-networks is defined by a supra connectivity matrix $T$:
\begin{equation}
T=
\begin{pmatrix}
L & P \\
P^{\top} & R
\end{pmatrix}
\end{equation}
Note that, $T$, $L$ and $R$ are intrinsically symmetric, meanwhile $P$ is non-symmetrical. Also note that in the supra connectivity matrix, nodes numbered from $i=1$ to $i=N_L$ belong to the left hemisphere, while nodes from $i=N_L +1$ to $i=N_L+N_R$ belong to the right one. The total number of nodes is therefore $N_T=N_L+N_R$. It is important to highlight that this methodology was applied for all frequency bands in both conditions. However, while the main body of this paper considers the $\alpha$ band in EC condition for its relevance in resting state phenomena, the Supp. Info. contains results for the remaining bands in both conditions.

\subsection*{Evaluating nodes and hemispheres centrality}
Eigenvector centrality correspond to the eigenvector $u_T$ associated to the first eigenvalue $\lambda_1$ of the network-of-networks $T$ as: $Tu_T=\lambda u_T$. 
Following the methodology of Aguirre et al\cite{Aguirre2013}, the centrality accumulated by the $L$ and $R$ hemispheres, $C_L$ and $C_R$, respectively, is obtained as the normalized sum of the centrality of the nodes belonging to each hemisphere, i.e, $C_L= \dfrac{\sum_{i=1}^{N_L}u_T(i)}{\sum_{i=1}^{N_T}u_T(i)}$ and $C_R= \dfrac{\sum_{i=N_L+1}^{N_L+N_R}u_T(i)}{\sum_{i=1}^{N_T}u_T(i)}$, with $u_T$ as the centrality over the whole network $T$. With such a normalization, the global centrality of $T$ is shared between both hemispheres following $C_L+C_R=1$. Intra-hemispherical eigenvector centralities $u_{L,R}(i)$ of the non-connected hemispheres can also be captured treating both sub-networks as whether they were isolated systems. Hence we can also categorize the importance of cortical regions (i.e., nodes) inside their own hemisphere (i.e., sub-network). Here, we call {\it hubs} those nodes with the highest intra-hemisphere centrality and they will be labeled as C nodes; while those nodes with lower intra-centrality are designed as {\it peripheral} nodes and labeled as P nodes.

\subsubsection*{The competition parameter $\Omega$}
In order to evaluate how the centrality depends on the connectivity between hemispheres, we consider the rewiring of the inter-hemispherical links, simply by reshuffling the $P$ matrix. By inspecting the consequences of rewiring $P$, we can gather the best strategies to optimize (maximize or minimize) the hemispherical centrality of a specific sub-network and, in turn, evaluate how far the actual centrality distribution is from any of the optimal cases.

The rewiring of $P$ is based on a deterministic search of the inter hemispherical links $p_{ij}$ that promotes the acquisition of centrality by one of the sub-networks forming a network-of-networks~\cite{Aguirre2013}. If we compute the largest eigenvalue $\lambda_1$ of the connectivity matrices of two networks (e.g., $L$ and $R$),  once we connect them, the network with higher $\lambda_1$ (suppose the dominant network is $L$) will be the one retaining more centrality (i.e., $C_L>C_R$). Nevertheless, there are two fundamental rules that can enhance/decrease the amount of centrality accumulated by a sub-network if the adequate connector nodes are selected. These rules that are explained in \cite{Aguirre2013} can be summarized as:
\begin{enumerate}
\item The network with higher $\lambda_1$ will accumulate more centrality if the connection with the other network is carried out through the peripheral nodes, i.e., peripheral-peripheral (PP) connection.
\item The network with lower $\lambda_1$ benefits from connecting to the other network through the central nodes, acquiring the highest possible centrality in this case (when compared with any other configuration), i.e. central-central (CC) connection.
\end{enumerate}

With these rules in mind, we analyzed the possible effects on the centrality by rewiring the inter-hemispheric sub-network $P$. The rewiring process is focused on obtaining the highest (lower) centrality of a given hemisphere, $L$ here for convenience. With this aim, we remove all connector links $p_{ij}$ from $T$ and re-incorporate them back in descending order, once at time. First, for the strongest $p_{ij}$ value, we computed an associated set of $(\frac{N}{2})^2$ values of $C_L$ by inserting $p_{ij}^{max}$ in all entries of the block matrix $P$. Once all positions of $P$ had been inspected we keep the weight such that we obtain the highest $C_L$. Then we repeat this procedure with the second highest value $p_{ij}$ taking into account the filled position of the previous $p_{ij}^{max}$, this time collecting a set of $(\frac{N}{2})^2-1$ values of $C_L$, and so forth up to finishing with that with the lowest $p_{ij}^{min}$. Once all weights $p_{ij}$ have been relocated, we estimate the $C_{max}^L$ ($C_{min}^L$) the global highest (lower) centrality accumulated by $L$ hemisphere, and conversely $C_{min}^R$ ($C_{max}^R$) the one accumulated by $R$ hemisphere, since $C_{max}^L+C_{min}^R=1$ ($C_{min}^L+C_{max}^R=1$).\\

When the configuration of inter-links associated to the highest and lowest centralities of $L$ hemisphere has been identified, we can define a competition parameter $\Omega_L$ that evaluates how the actual centrality distribution of $L$ is from the optimal configurations. It is important to highlight that the same procedure can be followed for $R$ hemisphere.

The parameter $\Omega_L=\Omega_L(C^L, C_{max}^L, C_{min}^L)$, is given by $\Omega_L=\frac{2(C^L-C_{min}^L)}{C_{max}^L-C_{min}^L}-1$.
Note that $\Omega_L$ is normalized as $-1\leq\Omega_L\leq1$. Values close to 1 (-1) indicate that the real distribution $C^L$ ($C^R=1-C^L$) is close to the optimal distribution for left (right) hemisphere. Importantly, when $\Omega_L$ is close to zero indicates that none of the hemispheres are close to their optimal configuration, in other words, they  are in a functional balance.

\subsection*{Robustness against node failure}
We characterized the network robustness by analyzing the impact of inter-hemispherical links on functional balance. For each subject, hemispherical sub-networks were thresholded by removing the lowest links' weights such that each sub-network remained connected by a single component. Both hemispherical sub-networks are then connected by adding the inter-hemispherical links once at a time in ascending order. We selected three configurations highlighting representative cases when hemispheres are either completely (Stage 1), relatively (Stage 2) and slightly connected (Stage 3) between them. For each of these stages, we carried out the analysis of the centrality robustness.

We define the {\it hemispherical local impact} $l^{L,R}_{imp}$, which accounts for the loss of centrality that a hemisphere suffers when one of its nodes is removed from the functional network, calculated (in the left hemisphere) as:
\begin{align}
& l_{imp}^L(i) = (\dfrac{C^L_*(i)-C^L}{C^L})*100\%
\end{align}
where $i$ is the node that has been removed, and $C^L$ and $C^L_*(i)$ are the centralities accumulated by the left hemisphere before and after the removal of node $i$, respectively. Since we can expect higher impacts when failures occur in the hemispherical hubs, we define the local contribution of a node as:
\begin{align}
& lc^L(i)=\frac{u^L_T(i)}{\sum_i^{N_L} u^L_T(i)}\times 100\%
\end{align}
which indicates the percentage of the hemispherical centrality that is captured by node $i$. The interplay between local impact and local contribution will be crucial to evaluate the robustness of the functional network under node failure. Local impact (contribution) for $R$ hemisphere are obtained in the same way, just replacing $L$ by $R$.


\section*{Acknowledgements}

This work was supported by the Spanish Ministry of Economy and Competitiveness (project No. FIS2013-41057-P). F. De Vico Fallani is supported by the French Agence Nationale de la Recherche (programme ANR-15-NEUC-0006-02). 

\section*{Author contributions statement}
F.D.V.F., J.M.B. and D.P. conceived the experiment,  F.D.V.F. conducted the experiment, J.M., M.C. and J.M.B. analyzed the results.  All authors wrote and reviewed the manuscript. 

\section*{Additional information}
\textbf{Competing financial interests} The authors declare no competing financial interests

\beginsupplement
\newpage
\section*{Supplementary Information}
The following results complement the information in the main text. Here we condensed the results obtained from frequency bands $\theta$, $\alpha$, $\beta$ and $\gamma$ for both EC and EO conditions.
\subsubsection*{Functional Balance in fully connected Network-of-Networks}
In Supplementary Fig. \ref{fig:s01}, we observe the positive trends between local and global centralities for the averaged connectivity matrices associated to each band and each condition (EC and EO). This positive trend suggests a sort of centrality balance when the full connected networks are taken into account regardless bands and conditions. All bands and condition behave in a similar way.
\begin{figure}[ht]
 \centering
 \includegraphics[width=0.65\textwidth]{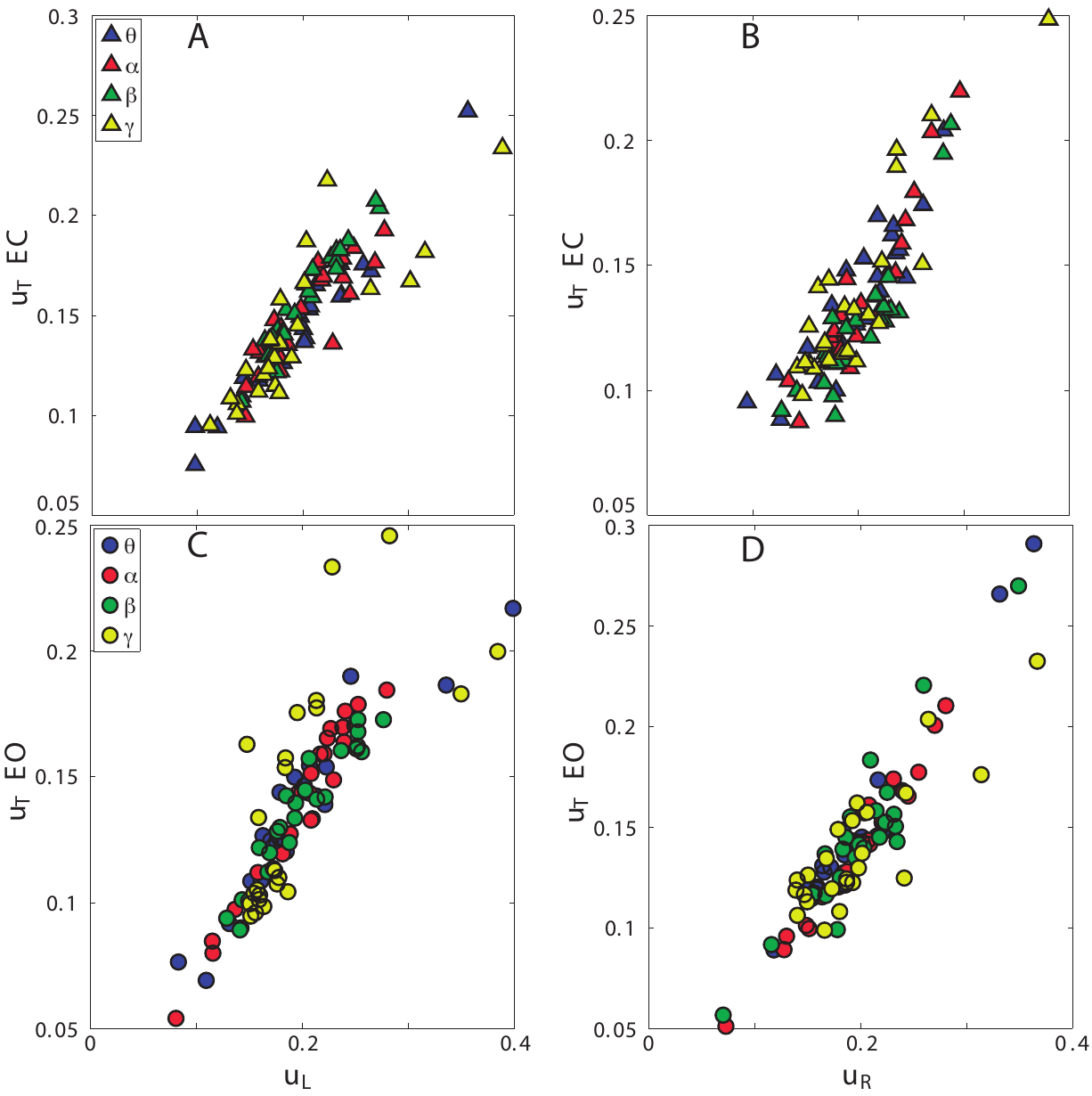}
 \caption{\textbf{Global and local centrality for all bands in EC and EO conditions.} Global centrality $u_T$ is obtained from the complete matrix $T$, when all functional connections between hemispheres are maintained (horizontal axes); while local centrality $u_{L,R}$ is extracted from the hemisphere matrices $L$ and $R$ when hemispheres are disconnected (vertical axes). Frequency bands $\theta$ (blue), $\alpha$ (red), $\beta$ (green), $\gamma$ (yellow) are drew in triangle and circle markers for EC and EO conditions, respectively.  Upper panel shows the EC condition for left (\textbf{A}) and right (\textbf{B}) hemispheres. Bottom panel: for left (C) and right (D) hemispheres in the EO condition.}\label{fig:s01}
\end{figure}

\begin{figure}[ht]
 \centering
 \includegraphics[width=1.0\textwidth]{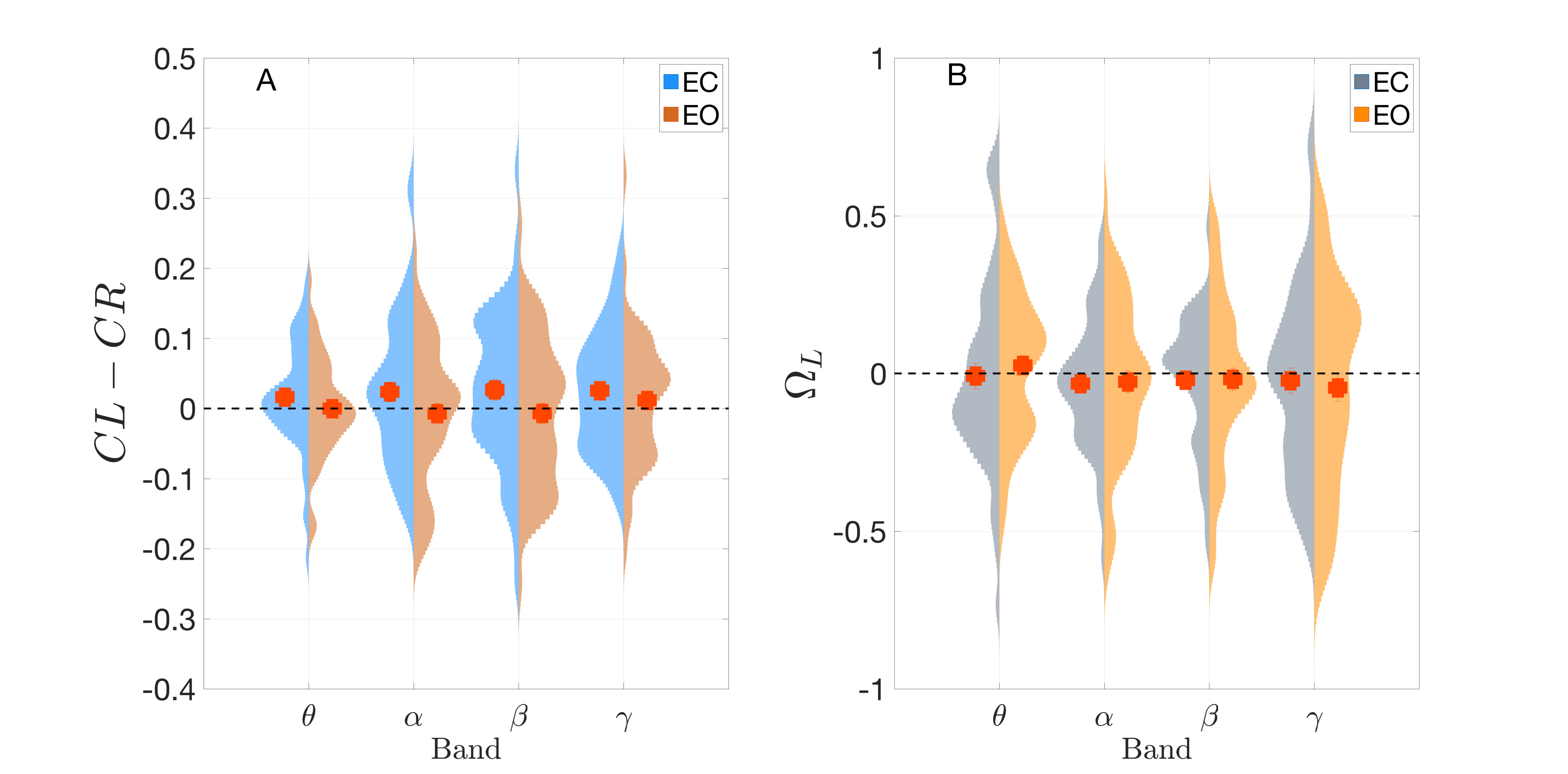}
 \caption{Centrality contrast and Competition parameter.
EC conditions (blue for centrality contrast and gray for $\Omega$) and EO (orange for centrality contrast and yellow for $\Omega$)) and four frequency bands: $\theta$, $\alpha$, $\beta$, $\gamma$.
\textbf{A}. Average values  of centrality contrast ($\langle C_L - C_R\rangle$) over 54 subjects are indicated by the red circles. \textbf{B}. Average of the Competition Parameter $\langle \Omega_L\rangle$ for all subjects.}\label{fig:s02}
\end{figure}

We also compute the centrality contrast and competition parameter for all bands in both conditions. Figure \ref{fig:s02} shows the violin plots and means of these global features. We do not find statistical differences between each mean values and a zero mean distribution by means of Mann-Whitney U Tests.

Summarized results are shown in Table \ref{tab:stab01}. Slight differences in the hemispherical importance are observed when both the EC and EO conditions are compared. We observe positive values of centrality contrast in the left hemisphere during the EC condition for all bands. On the contrary, we report negative values of centrality contrast when individuals open their eyes, which indicates a slight imbalance of centrality between the hemispheres in this condition (see Fig. \ref{fig:s02}\textbf{A} and second column of Table \ref{tab:stab01}).
\begin{table}[h]
\centering
\begin{tabular}{llcrlcrlcrlcrl}
\hline
\noalign{\smallskip}
& \multicolumn{3}{c}{$\mathbf{\langle C_L-C_R\rangle}$} & \multicolumn{3}{c}{$\langle\Omega_L\rangle$} \\

\noalign{\smallskip}
\cline{3-4}
\cline{6-7}
\noalign{\smallskip}
Band & & EC & EO & & EC & EO & \\
\noalign{\smallskip}
\bottomrule[1.2pt]

\noalign{\smallskip}
$\mathbf{\theta}$ && 
$0.016$&  $0.0$ && $-0.008$ &  $0.025$ & \\
\noalign{\smallskip}
$\mathbf{\alpha}$ && 
$0.023$ &  $-0.007$ && $-0.032$ &  $-0.026$ & \\
\noalign{\smallskip}
$\mathbf{\beta}$ && 
$0.026$ &  $-0.007$ && $-0.021$ &  $-0.018$ & \\
\noalign{\smallskip}
$\mathbf{\gamma}$ && 
$0.025$ &  $0.011$ && $-0.022$ &  $-0.044$ & \\
\hline
\end{tabular}
\caption{\textbf{Mean centralities contrast and competition parameter $\Omega_L$}. Numerical results obtained from Fig. \ref{fig:s02} in both conditions.}\label{tab:stab01}
\end{table}

Interestingly, results in Fig. \ref{fig:s02}\textbf{B} and Table \ref{tab:stab01}) show that $\langle \Omega_L \rangle$ for $\alpha$, $\beta$ and $\gamma$ are slightly negative. Since the competition parameter is defined taking the left hemisphere as reference, negative values of $\langle \Omega_L \rangle$ are consequence of an inter-hemispheric link distribution that slightly benefits the right hemisphere.

\newpage
\subsubsection*{Competition parameter: An example}
\begin{figure}[h]
 \centering
 \includegraphics[width=1.0\textwidth]{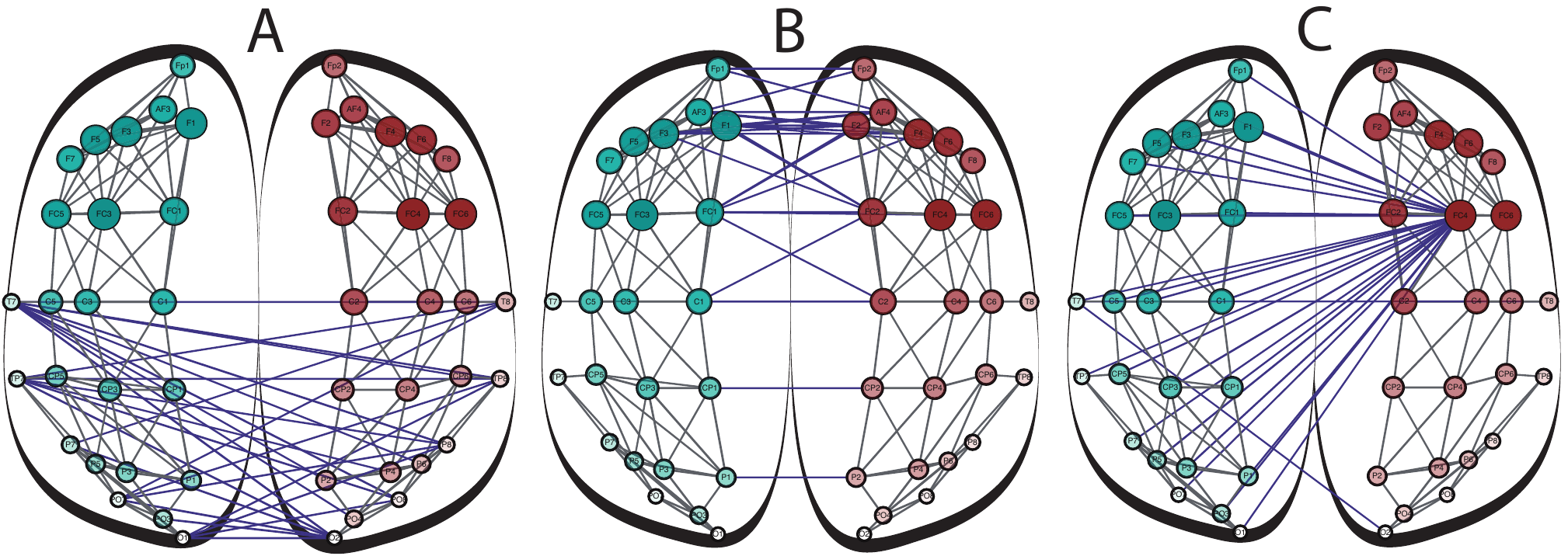}
 \caption{\textbf{Example of reshuffling inter-hemispherical links}. Three different configurations of the inter-hemispherical links. Intra-hemispherical connections are colored in grey, while blue is used for the inter-hemispherical links. Nodes' sizes and transparency are proportional to their local eigenvector centrality before connecting both hemispheres. The average $\alpha$-EC network has an arbitrary threshold that maintains the $16\%$ of the stronger links just for a better  visualization purpose. \textbf{A}. Peripheral nodes are connected (PP strategy), resulting in optimal strategy for increasing left hemisphere's centrality. \textbf{B}. Actual distribution of inter-hemispheric connections, which leads to a balance of the centrality distribution. \textbf{C}. Right hemisphere's optimal strategy is obtained by connecting central nodes (CC strategy).}\label{fig:s03}
\end{figure}

For illustrative purposes, Fig. \ref{fig:s03} exemplifies the rules explained in the Methods Section about the role of the connector nodes of each hemisphere. These connections lead to the best/worst centrality distribution for the average subject in alpha band. We show the configuration for inter-hemispheric links in three different cases for $\alpha$-EC.
The configuration that makes $L$ to acquire the highest centrality (Fig. \ref{fig:s03}\textbf{A}), relies on the peripheral-peripheral (PP) strategy. The real distribution of centrality (Fig. \ref{fig:s03}\textbf{B}) may use a mixture of CC (central-central), PP, CP and PC strategies. The configuration that allows $R$ hemisphere to get the highest centrality respect the real configuration (Fig. \ref{fig:s03}\textbf{C}) is based on the CC strategy. According to values in each strategy: PP leads to $C^L_{max}\approx0.7$, the actual configuration gives $C^L=0.49$ and CC grants $C^R_{max}\approx0.8$. We obtain $\Omega_L=-0.006$, which reveals the real inter-hemispheric connectivity pattern as promoter of the centrality balance between both hemispheres.

\newpage
\subsubsection*{Evolution of the left hemisphere centrality}
\begin{figure}[hbtp]
 \centering
 \includegraphics[width=1.0\textwidth]{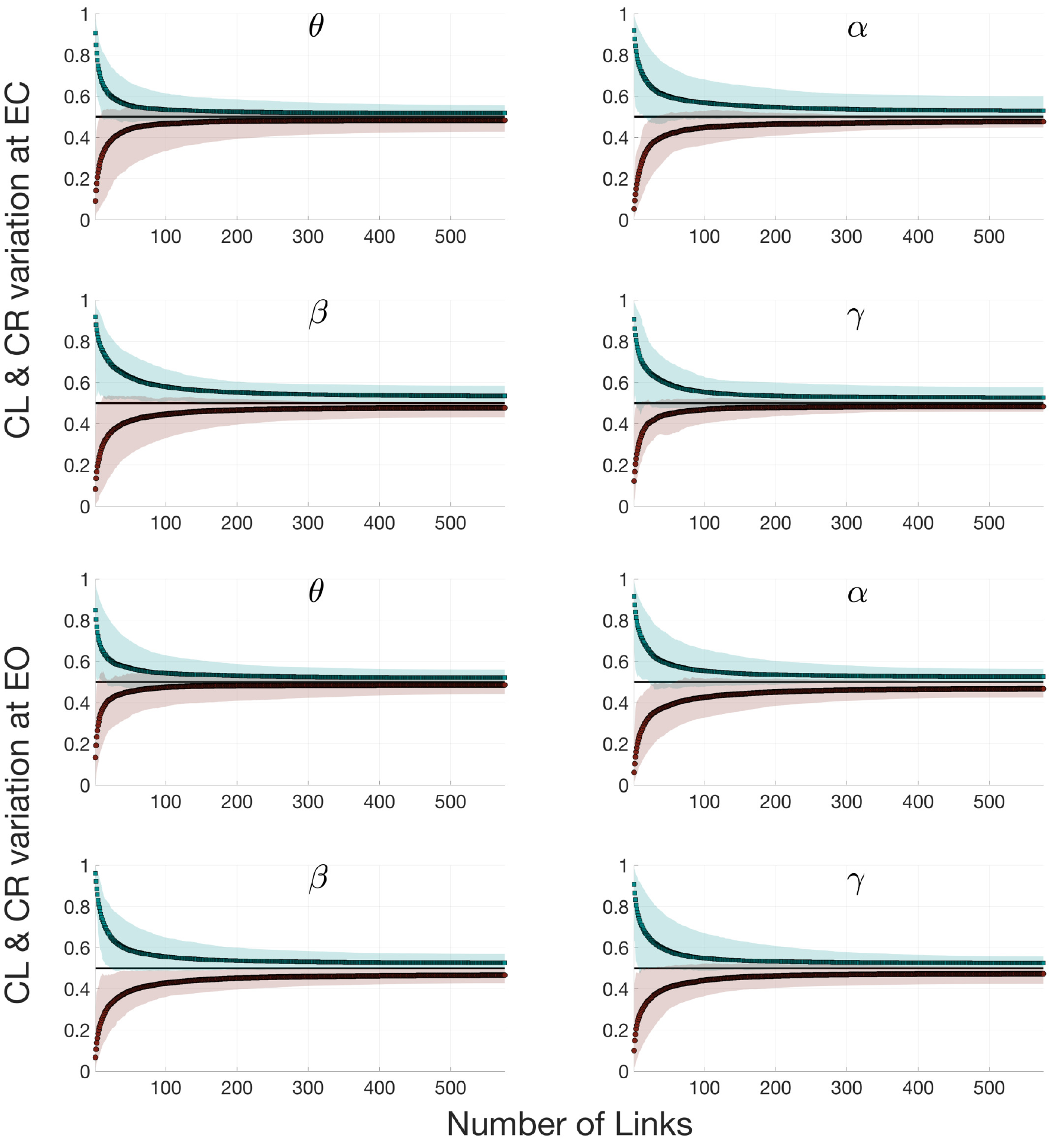}
 \caption{Hemispherical centrality $C_L$ vs. the number of inter-hemispherical links for all subjects. Each subplot shows a different combination of a condition (EC or EO) and frequency band ($\theta$, $\alpha$, $\beta$ and $\gamma$). Colors indicate whether the dominant hemisphere is the left (green) or the right (red). In each subplot, the region inside each curve includes the fifth and the 95th percentiles of $C_L$ and $C_R$ for all 54 subjects, with the average value plotted in dashed lines.}\label{fig:s04}
\end{figure}

Figure \ref{fig:s04} shows the evolution of $C_L$ (note that $C_R=1-C_L$) respect to the inter-hemispheric links. We distinguish between left- and right- dominant individuals based on the eigenvalue. The process of adding inter-hemispheric links shows a clear tendency:  the hemisphere that initially has the ``strongest" network (i.e., the higher $\lambda_1$) acquires a high amount of centrality when the number of inter-hemispherical links is low, but its centrality diminishes as the number of inter-links is increased. ``Weak" hemispheres behave just in the opposite way.

\newpage
\subsubsection*{Robustness and resilience of all bands}
\begin{figure}[hbtp]
 \centering
 \includegraphics[width=0.8\textwidth]{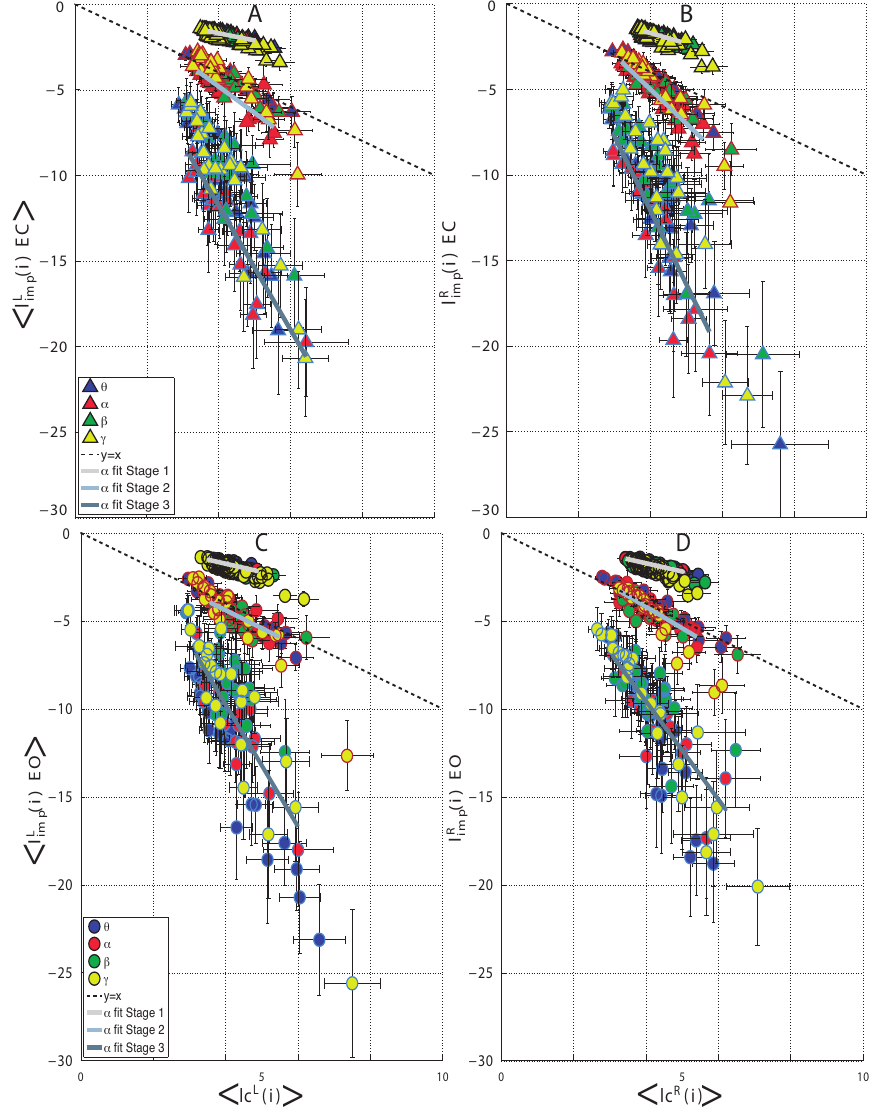}
 \caption{\textbf{Local impact $\langle l_{imp}^L(i)\rangle$ vs. local contribution $\langle lc^L(i)\rangle$ during a failure of node $i$}. In all plots, points (i.e., triangles/circles) correspond to single-node failures (i.e., a single node is removed from the functional network). Triangles (circles) of the Upper (bottom) plots refer to the EC (EO) condition. Left (right) panel refers to the  $L$($R$) hemisphere. In each subplot three groups can be identified: (1) Stage 1 corresponds to the upper cloud of nodes, (2) Stage 2 to the middlle cloud and (3) Stage 3 to the bottom group of nodes. The dashed line corresponds to $\langle l_{imp}^L(i)\rangle=\langle lc^L(i)\rangle$. For each stage, a solid line allows to better follow the linear correlation between the local contribution and the local impact.}\label{fig:s05}
\end{figure}

We report the results of local impact vs local contribution in the three stages for all bands in both conditions. Tables with the slopes of linear fits between the local impact and local contribution are presented in Table \ref{tab:tab03s} and \ref{tab:tab04s}. In Fig. \ref{fig:s05} network damage is in the vertical axes and local contribution is in horizontal axes. The values of the three stages are shown for all bands and conditions.

\begin{table}[h]
\centering
\begin{tabular}{llcrlcrlcrlcrl}
\hline
\noalign{\smallskip}
& \multicolumn{3}{c}{Stage 3} & \multicolumn{3}{c}{Stage 2} & \multicolumn{3}{c}{Stage 1} \\

\noalign{\smallskip}
\cline{3-4}
\cline{6-7}
\cline{9-10}
\noalign{\smallskip}
Band & & $m_{EC_L}$ & $m_{EO_L}$ & & $m_{EC_L}$ & $m_{EO_L}$ & & $m_{EC_L}$ & $m_{EO_L}$ & \\
\noalign{\smallskip}
\bottomrule[1.2pt]

\noalign{\smallskip}
$\mathbf{\theta}$ && 
$-3.784$ &  $-4.296$ && $-1.238$ &  $-1.357$ && $-0.479$&  $-0.537$ & \\
\noalign{\smallskip}
$\mathbf{\alpha}$ && 
$-3.629$ &  $-3.466$ && $-1.541$ &  $-1.021$ && $-0.421$ &  $-0.465$ & \\
\noalign{\smallskip}
$\mathbf{\beta}$ && 
$-3.107$ &  $-2.416$ && $-1.262$ &  $-1.082$ && $-0.599$ &  $-0.631$ & \\
\noalign{\smallskip}
$\mathbf{\gamma}$ && 
$-4.034$ &  $-4.098$ && $-1.857$ &  $-2.013$ && $-0.686$ &  $-0.748$ & \\
\hline
\end{tabular}
\caption{\textbf{Correlation between the local impact and the local contribution (left hemisphere)}. Slopes of linear fits of $\langle l_{imp}^L\rangle$ vs. $\langle l_c^L\rangle$ for all bands, conditions and stages associated to Fig. \ref{fig:s05} \textbf{A} y \textbf{C}. Note the increase of the slope from Stage 1 to Stage 3 as that described in the main text of the manuscript.}\label{tab:tab03s}
\end{table}

\begin{table}[h]
\centering
\begin{tabular}{llcrlcrlcrlcrl}
\hline
\noalign{\smallskip}
& \multicolumn{3}{c}{Stage 3} & \multicolumn{3}{c}{Stage 2} & \multicolumn{3}{c}{Stage 1} \\

\noalign{\smallskip}
\cline{3-4}
\cline{6-7}
\cline{9-10}
\noalign{\smallskip}
Band & & $m_{EC_R}$ & $m_{EO_R}$ & & $m_{EC_R}$ & $m_{EO_R}$ & & $m_{EC_R}$ & $m_{EO_R}$ & \\
\noalign{\smallskip}
\bottomrule[1.2pt]

\noalign{\smallskip}
$\mathbf{\theta}$ && 
$-3.942$ &  $-4.660$ && $-1.759$ &  $-1.070$ && $-0.623$ &  $-0.441$ & \\
\noalign{\smallskip}
$\mathbf{\alpha}$ && 
$-4.322$ &  $-2.788$ && $-2.000$ &  $-1.222$ && $-0.608$ &  $-0.466$ & \\
\noalign{\smallskip}
$\mathbf{\beta}$ && 
$-3.118$ &  $-1.430$ && $-1.653$ &  $-1.064$ && $-0.532$ &  $-0.586$ & \\
\noalign{\smallskip}
$\mathbf{\gamma}$ && 
$-4.480$ &  $-3.534$ && $-2.462$ &  $-2.240$ && $-0.833$ &  $-0.933$ & \\
\hline
\end{tabular}
\caption{\textbf{Correlation between the local impact and the local contribution (right hemisphere)}. Slopes of the linear fits of $\langle l_{imp}^R\rangle$ vs $\langle l_c^R\rangle$ for all bands, conditions and stages associated to Fig. \ref{fig:s05} \textbf{C} y \textbf{D}. Note the increase of the slope from Stage 1 to Stage 3 as that described in the main text of the manuscript.}\label{tab:tab04s}
\end{table}

\begin{figure}[ht]
 \centering
 \includegraphics[width=1.0\textwidth]{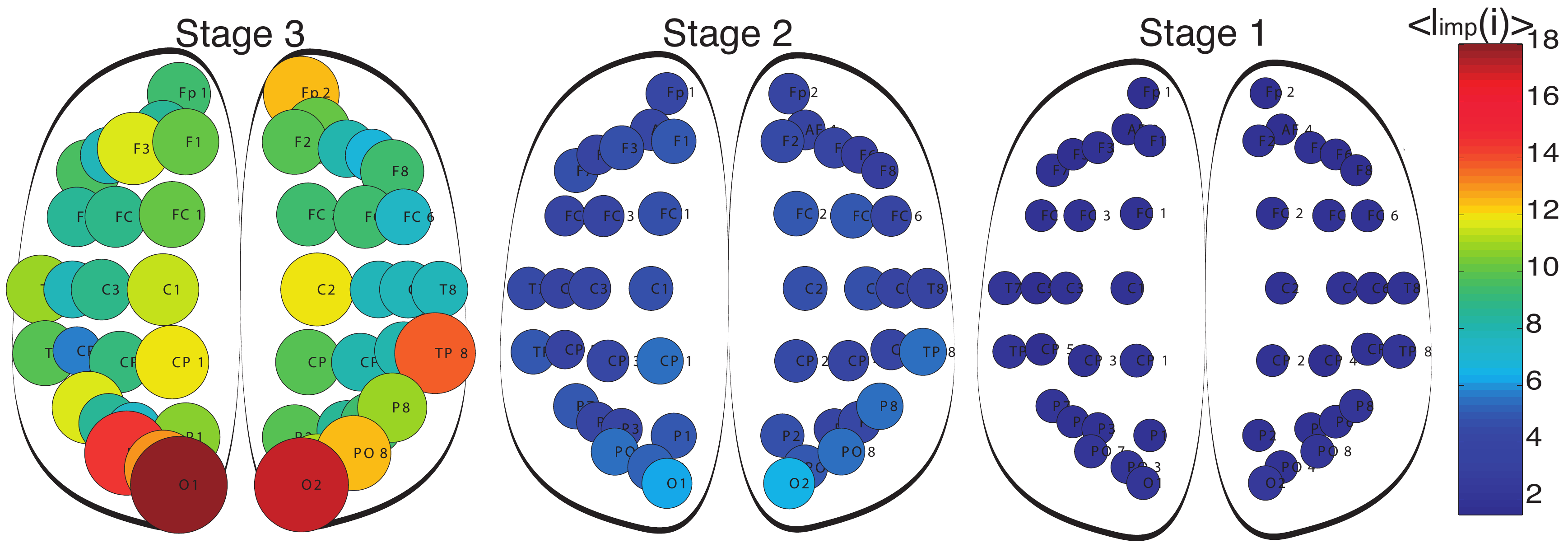}
 \caption{\textbf{Local impact in the $\alpha$ band for EO condition}. 
Average local impacts in $\alpha$ band and EO condition. The radius of each node is proportional to the average local impact. Note how the local impact of the nodes increases as we move from Stage 1 to Stage 3 indicating the occipital lobe as the most vulnerable region.}\label{fig:s06}
\end{figure}

\newpage
Figure \ref{fig:s06} shows the topological distributions of averaged local impact nodes' positions ate three different stages. In this case, we focus only in $\alpha$ band.

\subsubsection*{Local impact on centrality, clustering and shortest path}
\begin{figure}[ht]
 \centering
 \includegraphics[width=0.6\textwidth]{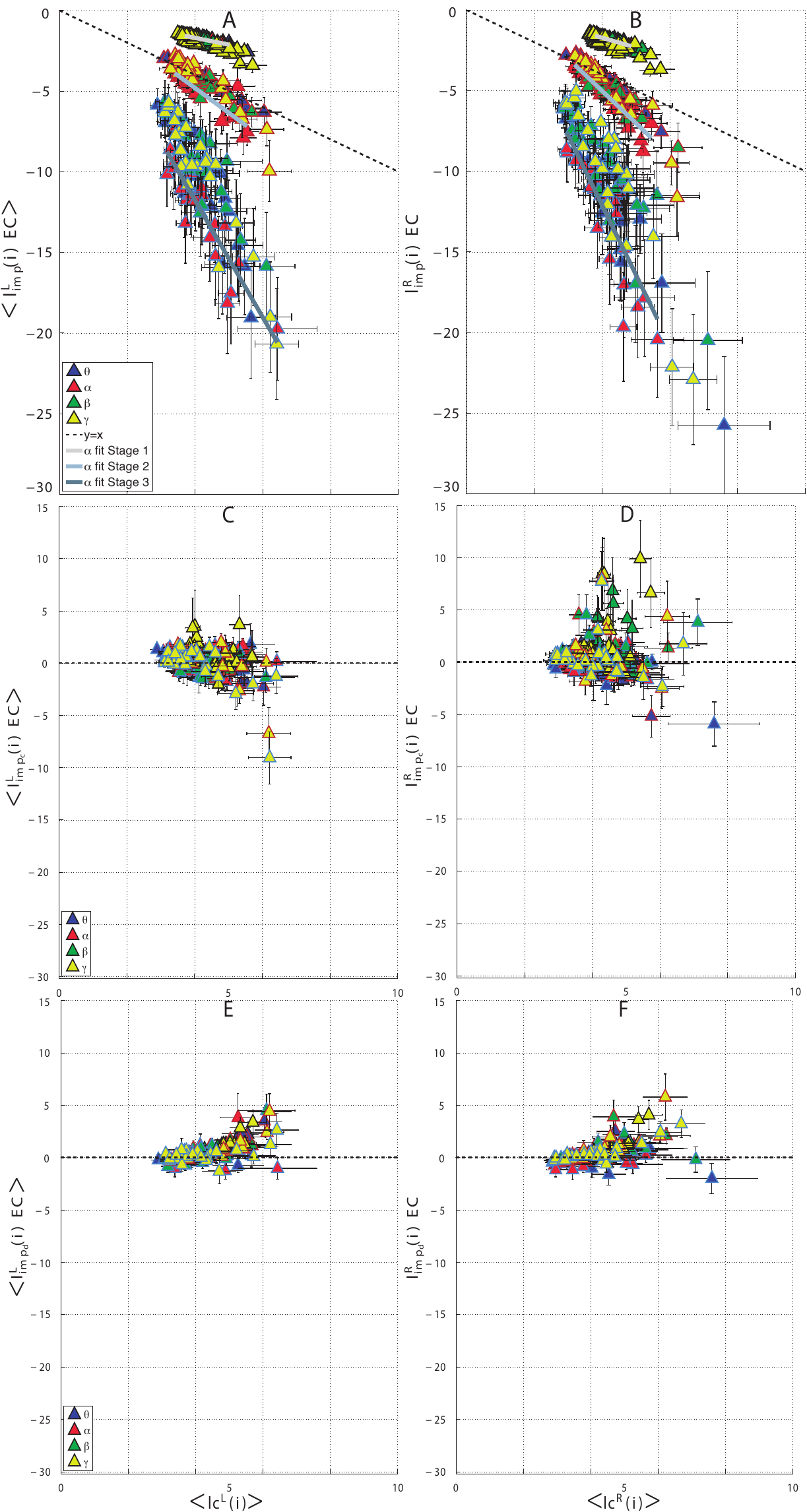}
 \caption{\textbf{Local impact for centrality, clustering and shortest path. EC condition}. 
Upper panels: Local impact on the hemispherical centrality for left (\textbf{A}), and right (\textbf{B}) hemispheres. Middle panels: Local impact on the clustering coefficient for the left (\textbf{C}) and right (\textbf{D}) hemispheres. Bottom panel: Local impact on the shortest path for left (\textbf{E}) and right (F) hemispheres.}\label{fig:s07}
\end{figure}

\begin{figure}[ht]
 \centering
 \includegraphics[width=0.65\textwidth]{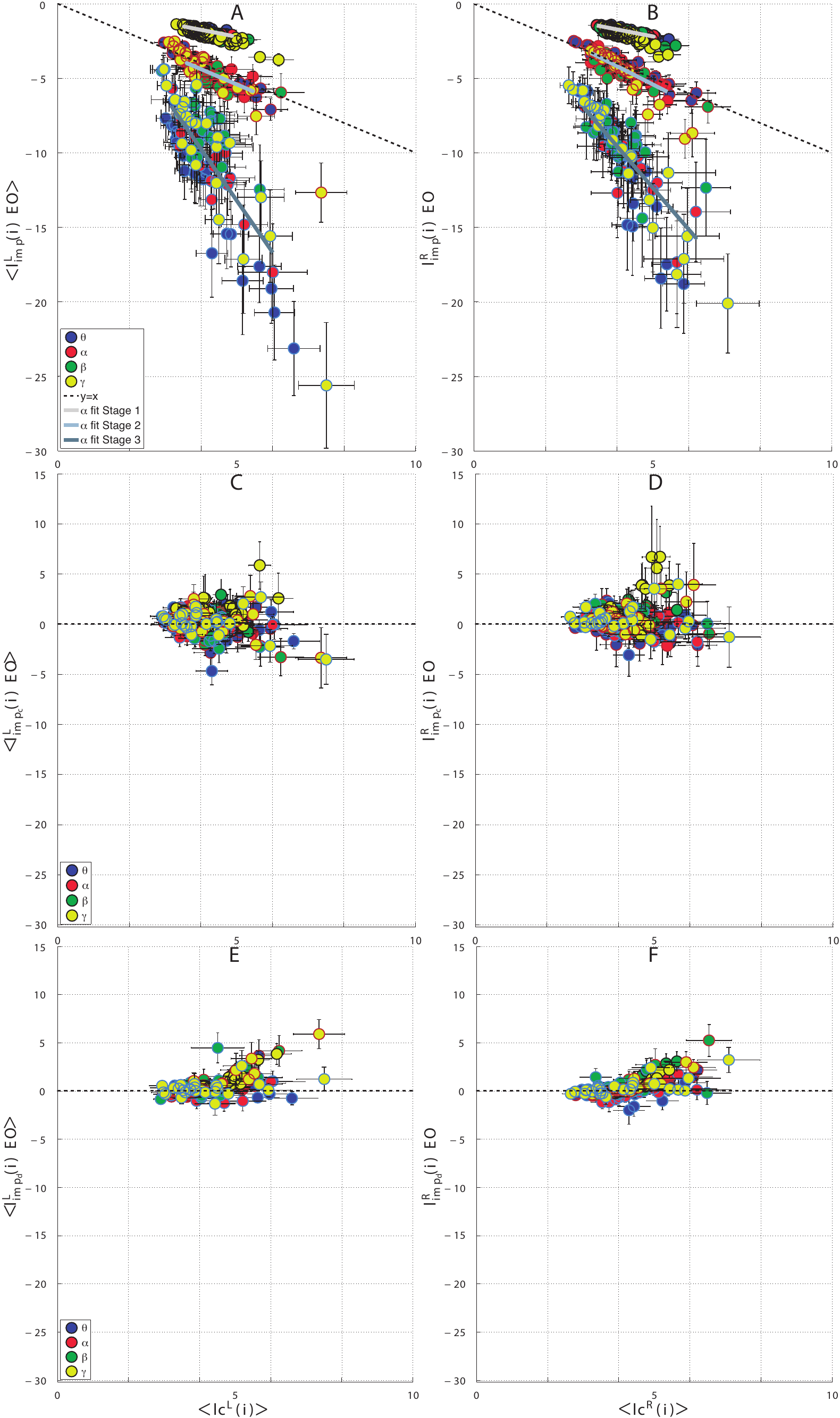}
 \caption{\textbf{Local impact for centrality, clustering and shortest path. EO condition}. 
Upper panels: Local impact on hemispherical centrality for the left (\textbf{A}), and right (\textbf{B}) hemispheres. Middle panels: Local impact on the clustering coefficient for the left (\textbf{C}) and right (\textbf{D}) hemispheres. Bottom panel: Local impact on the shortest path for the left (\textbf{E}) and right (F) hemispheres.}\label{fig:s08}
\end{figure}

Figures \ref{fig:s07} and Fig. \ref{fig:s08} show the behaviour of local impact on centrality,  clustering ($\langle l_{imp_c}(i)\rangle$) and shortest path ($\langle l_{imp_d}(i)\rangle$) respect to the local contribution $\langle l_c(i)\rangle$ for all bands and both conditions. In both plots each condition, EC and EO, is represented with triangles and circles, respectively. Frequency bands are also differentiated: $\theta$ (blue), $\alpha$ (red), $\beta$ (green) and $\gamma$ (yellow).
In contrast to the local impact on centrality, clustering and shortest path do not show important changes, no matter the stage considered.

\end{document}